# Improving Access to Trade and Investment Information in Thailand through Intelligent Document Retrieval

Sirinda Palahan

University of the Thai Chamber of Commerce, Bangkok 10400, Thailand

Email: sirinda_pal@utcc.ac.th.

## Abstract

Overseas investment and trade can be daunting for beginners due to the vast amount of complex information. This paper presents a chatbot system that integrates natural language processing and information retrieval techniques, aiming to simplify the document retrieval process. The proposed system identifies the most relevant content, enabling users to navigate the intricate landscape of foreign trade and investment more efficiently. Our methodology combines the BM25 model and a deep learning model to rank and retrieve documents aiming to reduce noise in the document content, thereby enhancing the accuracy of the results. Experiments with Thai natural language queries have proven the effectiveness of our system in retrieving pertinent documents. A user satisfaction survey further validated the system's effectiveness. Most respondents found the system helpful and agreed with the suggested documents, indicating its potential as a beneficial tool for Thai entrepreneurs navigating foreign trade and investment.

## 1   Introduction

Foreign trade and investment can be a daunting experience for beginners due to the many rules, regulations, and customs involved. The internet has become a valuable resource for finding relevant documents, such as articles and guides that cover all aspects of foreign trade and investment. However, with the growing number of documents available, it is becoming increasingly difficult to find useful information, and beginners can feel overwhelmed and intimidated by the comprehensive nature of the materials.

To address this issue, traditional information retrieval methods such as TF-IDF, a term-based approach, have been utilized to rank documents based on their relevance to a user's query. Two components of documents that can be considered for retrieval are the document content and headings. The majority of research studies (Wu, Luk, Wong, & Kwok, 2008; Guo, Alamudun, & Hammond, 2016; Zaragoza, Craswell, Taylor, Saria, & Robertson, 2004; Robertson, Zaragoza, et al., 2009; Robertson, Zaragoza, & Taylor, 2004; Zhai & Lafferty, 2017; Hiemstra, 2001; Amati & Van Rijsbergen, 2002) employ the content of documents in term-based approaches. Retrieving documents based on their entire content can yield more comprehensive outcomes, as it considers all the information encompassed within the document, including details that might not be expressed in the headings. However, the full content of a document may contain a significant amount of irrelevant or extraneous information, which can make it difficult to accurately retrieve relevant results. Moreover, the accuracy of such an approach can be limited by the quality of the tokenization process. If the tokenizer fails to accurately represent the terms in the document, the results may be inaccurate, leading to less relevant search results. The document headings can provide valuable information for accurate document retrieval. Retrieving documents based on the headings can avoid noise in the content that might be irrelevant to the query. It can be faster than retrieving documents based on their full content. This is because headings are often shorter and more focused, making it easier to quickly determine their relevance to the query. Our proposed method addresses this issue by considering both the document content and headings for retrieval, resulting in more accurate and efficient search results.

To enhance document retrieval accuracy, deep learning models have been integrated into the system to consider not only the terms but also the semantic aspects of documents. These models employ techniques such as neural embeddings and deep neural networks to learn distributed representations of text, which can capture the underlying semantic meaning of words in a document. However, utilizing deep learning models for low-resource languages like Thai can be particularly challenging due to the complexity of the language. Thai lacks clear word boundaries, exhibits sentence pattern variations, and contains compound word ambiguity, all of which contribute to its complexity and make the development of effective IR techniques difficult. Nevertheless, we believe that leveraging natural language processing techniques is critical to retrieving relevant documents in Thai to address entrepreneurs' queries. This paper offers two primary contributions. First, we propose a deep-learning-based model specifically designed to interpret user queries in Thai and retrieve the most pertinent documents. Second, we integrate this model into a chatbot, providing a comprehensive end-to-end solution tailored to the business and economic sectors. By doing so, the model is tested in a practical environment, ensuring its utility and effectiveness. The ultimate goal of this system is to deliver succinct and relevant document content to aid Thai entrepreneurs in maneuvering through foreign trade and investment processes. This paper is organized as follows: Section 2 provides an overview of related works, while Section 3 discusses the methodology of the proposed system. Section 4 presents the experimental results, and Section 5 provides conclusions and future directions for this work.

## 2   Literature review

### 2.1   Information retrieval

Several studies have been conducted in the Information Retrieval community on ad-hoc document retrieval tasks. In the ad hoc retrieval task, a system is given a user query, and its task is to return the top most relevant documents in a corpus, according to some ranking functions. The traditional Information Retrieval approaches evaluate users' queries primarily based on a word match at a syntactic level (Wu et al., 2008; Guo et al., 2016; Zaragoza et al., 2004; Robertson et al., 2009, 2004; Zhai & Lafferty, 2017; Hiemstra, 2001; Amati & Van Rijsbergen, 2002). The TF-IDF approach considers the frequency of words within a document, i.e., term-frequency (TF), and the inverse document frequency (IDF) of the same words across the entire collection of documents. While the TF-IDF approaches have been widely used in information retrieval, they have limitations in accurately ranking documents based on their relevance to a user's query. Because they rely solely on the occurrence of query terms within the document, without considering other factors that may affect the relevance of the document to the user's query, such as document length or the proximity of query terms within the document. To address these limitations, the BM25 approach was developed, which takes into account not only the frequency of terms in a document but also the length of the document and the average length of documents in the collection. By incorporating these additional factors, BM25 has been shown to outperform TF-IDF and language modeling approaches in various retrieval tasks (Naseri, Dalton, Yates, & Allan, 2021; Soni & Roberts, 2020; Lan, Ge, & Kong, 2019)

However, both approaches do not consider the order and semantics of words within the document. The advancement of deep learning methods has led to an improvement in IR tasks and ad-hoc information retrieval tasks. The Deep Learning (DL) methods have been employed in IR to add semantic aspects to the retrieval tasks. There are three broad categories of DL approaches to IR based on their influence on the query representation, the document representation, or ranking models (Mitra, Craswell, et al., 2018). Mitra et al. (Mitra, Nalisnick, Craswell, & Caruana, 2016) proposed the Dual Embedding Space Model (DESM) to map the query words into the input space and the document words into the output space and calculate a relevance score by combining the cosine similarities across all pairs of the query-document words. The experiments showed that the proposed model could rank top documents better than a traditional ranking method based on TF-IDF. Yang et al. (Yang, Zhang, & Lin, 2019) applied BERT, a deep learning model, to improve the document representation where documents are longer than the length of input BERT. The authors overcome the problem by applying inference over individual sentences and combining sentence scores into document scores. The approach was tested on TREC microblog and newswire test collections, and the test results showed their approach was effective. One of the first ranking models is the Deep Structured Semantic Model (DSSM) (Huang et al., 2013), a neural ranking model for the ad-hoc retrieval task. In this approach, queries and documents are mapped into a low-dimensional space, and relevance is calculated by determining the distance between them. Finally, Lu   Li (Lu & Li, 2013) applied a deep learning model to matching tasks, such as finding relevant answers to a givenquestion. The authors compared their deep matching model to the inner-product-based model. The empirical results showed that the proposed model outperformed inner-product-based matching models on real-world datasets.

There has been an increasing interest in combining traditional term-based approaches with semantic approachesin information retrieval, aiming to capture the benefits of both approaches (Galke, Saleh, & Scherp, 2017; Noraset, Lowphansirikul, & Tuarob, 2021). This hybrid approach leverages the strengths of term-based approaches in capturing local and discrete representations of text, as well as the strengths of deep learning-based approaches in capturing the semantic meaning of words and phrases. Mitra et al. (Mitra, Diaz, & Craswell, 2017) proposed a document ranking model that combines traditional methods relying on terms in the body text and newer models relying on distributed representations using two separate deep neural networks. One network matches the query and document using local representation, and the other matches them using distributed representation. The two networks are trained together as a single neural network, and the authors demonstrated that this approach outperforms traditional methods and other models based solely on neural networks. Galke et al. (Galke et al., 2017) evaluated different techniques that use word embedding for information retrieval in short query scenarios. The techniques include word centroid similarity, paragraph vectors, Word Mover's distance, and IDF re-weighted word centroid similarity. The results show that word centroid similarity is the best technique, especially when the word frequencies are re-weighted using IDF before aggregating the respective word vectors of the embedding. In some cases, the proposed cosine similarity of IDF re-weighted word vectors outperforms the traditional TF-IDF baseline by 15%.

## 2.2   Natural Language Processing for the Thai Language

Natural language processing (NLP) is a branch of computer science, specifically artificial intelligence, which aims to train computers to understand spoken words and written text, much like a human would. As NLP research has expanded into more languages over the past decade, the number of languages covered has also increased. Similarly, NLP development for the Thai language has been growing continuously. The combination of natural language processing (NLP) and information retrieval has enabled researchers to develop systems capable of retrieving articles and answering questions automatically. Limkonchotiwat (Limkonchotiwat et al., 2021) proposed a novel system called WabiQA, which leverages NLP and information retrieval techniques to answer questions in the Thai language. Specifically, the system uses Thai Wikipedia articles as the knowledge source and retrieves the article most likely to contain the answer. A bidirectional LSTM model is then employed to locate candidate answers, which are ranked and presented to the user. The system was evaluated in the National Software Contest 2019 and outperformed competitors' systems by a significant margin. The research findings suggest the potential to develop intelligent NLP applications for low-resource languages such as Thai using existing tools and resources, with implications for a wide range of NLP tasks.

In Kawtrakul et al. (Kawtrakul et al., 2000), a framework was introduced to develop a Thai document retrieval system using NLP techniques. The researchers employed a rule-based strategy to extract phrases and identify the relations between the terms in the phrases. These phrases were then used to create multilevel indexes, which were used to retrieve documents with inverted indexes that match the input query and belong to the same category. The experimental findings indicated that using NLP techniques to create an index set was more effective than not using them. In Sukhahuta Smith (Sukhahuta & Smith, 2000), an approach was proposed to enhance the presentation of Thai language documents by utilizing syntactic analysis. The approach involved analyzing documents and transforming them into a syntactic tree structure. The construction of the syntactic tree followed a specific procedure: first, the document was tokenized using predefined rules for a particular lexicon; next, each word was assigned a part-of-speech tag. Then, the context-free phrase structure grammar was applied to identify the syntactic surface structure, and finally, a syntactic tree was generated from the tagged words. The resulting tree was used to extract a list of concept definitions for each document. The evaluation of the proposed approach was based on the accuracy of the extracted concepts. The precision and recall were measured, and the average values obtained were 42% and 70%, respectively.

### 3 Methodology

This section describes the architecture of our system and its process for suggesting relevant documents to users based on their queries. As depicted in Figure 1, the system is composed of four key modules: a conversational interface, a document processing module, a query processing module, and a ranking module. The document processing module initially preprocesses the documents and stores document representations in a database for efficient retrieval. When a user submits a query via the conversational interface, the query processing module steps in to extract the keywords from the query, subsequently utilizing these keywords to represent the query. Following this, the ranking module evaluates the documents based on their relevance to the user's query. The document earning the highest relevance score is selected and recommended as the most suitable response to the user's query. It's important to note that the document processing, query processing, and ranking modules were implemented on a cloud service, ensuring scalability and accessibility. Additionally, an API gateway is employed to connect the system to the conversational interface, allowing for seamless interaction between the user and the system. The subsequent subsections provide a detailed breakdown of each component and its role in the overall operation of the system.

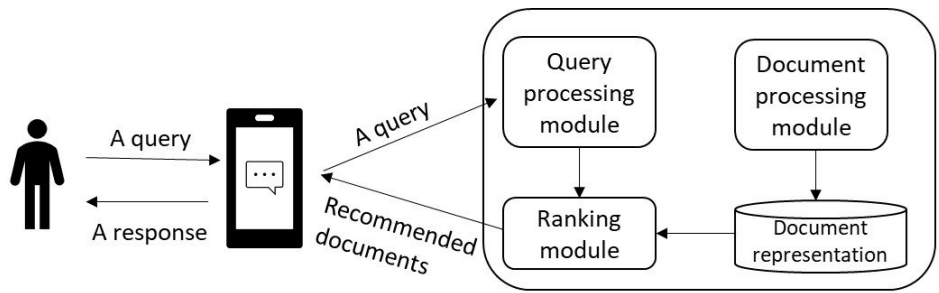

Figure 1: The architecture of the system.

## 3.1 The Conversational Interface

The first component of the system is a conversational interface acting as a chatbot to interact with users. In this work, the LINE messenger app was chosen as the conversational interface because it is Thailand's most-used messaging app (NAVER Corporation, 2020). This choice ensures that the system is easily accessible to Thai users, leveraging the popularity and familiarity of the LINE app in the country.

The conversational interface operates within the LINE messenger app and is showcased in Figure 2: (Left) Conversational interface showing query response; (Right) Document view when opened. When a user submits a query, such as "ขอข้อมูลเกี่ยวกับการนำเข้าสินค้าในเมียนมาร์หน่อยค่ะ" (Please provide information about importing goods into Myanmar), the system initiates the document retrieval process. It then identifies and delivers the top three most relevant documents directly to the user through the LINE messenger interface, in the form of summaries and direct links to the full documents, as shown in Figure 2 (Left). Users can then click on these links to read the full documents, as shown in Figure 2 (Right). This approach ensures a quick and streamlined response to the user's information needs.

## 3.2 Document Processing Module

This study leverages guidebooks on foreign investment and trade as a rich source of information to answer entrepreneurs' queries. These guidebooks, however, encompass various topics, some of which may not be relevant to a particular query. To mitigate this issue, we divided the guidebooks into smaller, more manageable sections based on their headings. This approach ensures that each returned document contains focused and concise information. In our methodology, headings are treated as summaries of their respective sections and are processed as follows. Initially, we analyzed the guidebooks, which were downloaded in PDF format. We employed regular expressions to identify the start and end points of the table of contents, from which we extracted the headings and their corresponding page numbers. Each heading's content was then isolated into individual PDF files by extracting the specific pages under each heading. The headings were further utilized as file names to facilitate document retrieval. Subsequently, the system processes the headings, tokenizing sentences into words and extracting keywords. In the keyword extraction process, our system leverages the ORCHID Part-of-Speech (POS) tagging system, which is specifically designed for the Thai language.

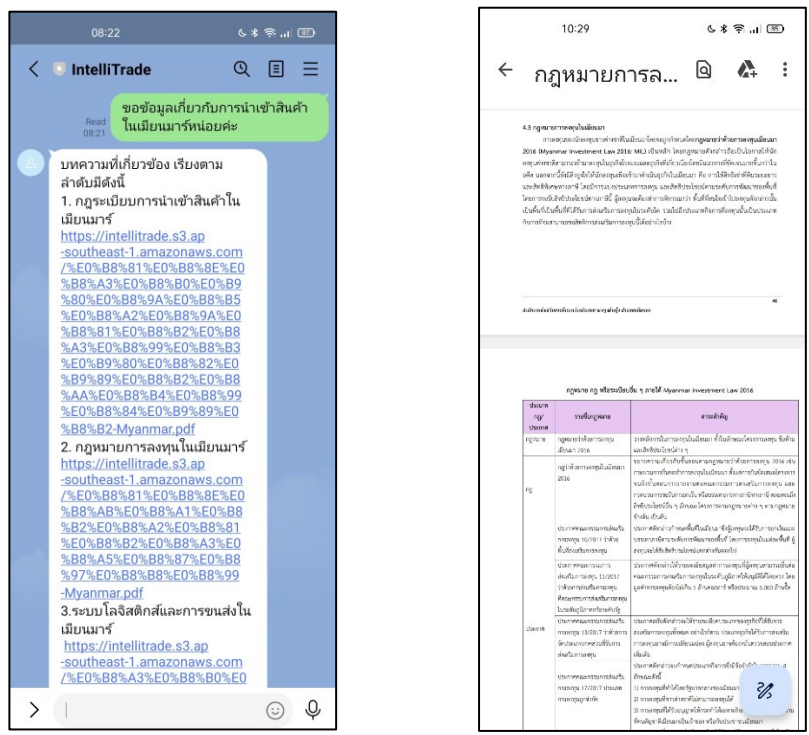

Figure 2: (Left) Conversational interface showing query response; (Right) Document view when opened.

In the keyword extraction process, our system leverages the ORCHID Part-of-Speech (POS) tagging system, a specialized resource designed for processing the Thai language. The ORCHID POS tagging system classifies individual words into their corresponding grammatical categories such as nouns, verbs, and adjectives, and annotates these tags to the words. This annotation allows for a more accurate and contextual interpretation of Thai natural language queries. The tagged and annotated words are essential for identifying relevant keywords with greater precision. However, not all tags are considered for keyword extraction. We focus on a select set of tags that are most likely to include meaningful keywords. The selected tags are as follows:

- CMTR: Measurement classifier - This tag is used for words that classify measurements, a common

feature in the Thai language that often carries significant semantic weight.

- NPRP: Proper noun - Proper nouns often include crucial information such as names of places, organizations, or individuals.

- NCMN: Common noun - Common nouns are often key to understanding the main topics covered in a document

- NTTL: Title noun - Title nouns, which are used to denote titles or positions, can provide valuable context.

- VACT: Active verb - Active verbs describe actions and are often crucial for understanding the activities discussed in a document.

- VSTA: Stative verb - Stative verbs express states or conditions and can contribute valuable information about the circumstances or characteristics described in a document.

These specific tags are selected based on their likelihood to contain meaningful and relevant information. Once the keywords are extracted, they are then mapped to word vectors using the Large Thai Word2Vec model (Phatthiyaphaibun, 2022). This pre-trained deep learning model, specifically designed for the Thai language, employs the Word2Vec architecture. It has been trained on an extensive corpus of Thai text, capturing the semantic nuances of Thai words and phrases. The model creates a vector representation of each word by analyzing its contextual relationship with surrounding words in the corpus. To create a vector representation for each document, the system calculates the mean of all the keyword vectors. This strategy provides a condensed yet comprehensive representation of each document, which effectively accounts for the complexities and nuances of the Thai language.

| Heading | Keywords |
|---|---|
| ลักษณะภูมิประเทศ (Characteristic of terrain) | ['ลักษณะ', 'ภูมิประเทศ'] [ 'characteristic', 'terrain'] |
| สภาพภูมิอากาศ (Characteristic of climate) | ['ลักษณะ', 'ภูมิอากาศ'] [ 'characteristic', 'climate[ |
| วัฒนธรรมและมารยาททางธุรกิจ (Business culture and etiquette) | ['วัฒนธรรม', 'มารยาท', 'ธุรกิจ'] [ 'culture', 'etiquette', 'business'] |

Table 1: The original headings and corresponding keywords

To illustrate, Table 1 shows the original headings and their corresponding keywords in Thai, along with English translations in parentheses.

Moreover, to represent each document more comprehensively, our approach incorporates the content of each document alongside the extracted heading keywords. This involves tokenizing the content into sentences and subsequently into words. Given that the conversion from PDF to text files can sometimes result in incorrect words, especially in Thai due to its complex script and lack of spaces between words, our system employs Peter Norvig's spelling correction algorithm (Norvig, 2016) to maintain the accuracy of the content. Next, stop words are removed, an essential step in reducing noise and focusing on meaningful words, especially important given the high context nature of the Thai language. The resulting tokens, along with the extracted keywords, are then used to represent the document.

In summary, the document representation includes two vectors: one containing the heading keywords and the other containing the content tokens. The system preprocesses and stores heading keywords and content tokens in a database, which are then retrieved for ranking when a query is received. This enables efficient and quick retrieval of relevant documents based on the input query.

### 3.3   Query Processing Module

Processing user queries indeed presents a greater complexity compared to handling document headings and content. This complexity primarily arises due to the conversational nature of queries, which can often include stop words, misspelled words, or unnecessary characters. Furthermore, nuances inherent to the Thai language—such as its lack of space between words, multiple valid spellings for the same word, and the use of tone marks—add layers of difficulty to the processing of user queries. In response to these challenges, our system executes several steps to process user queries effectively. The first step is tokenization, which splits the user's query into individual words. Tokenization is particularly challenging in Thai due to the absence of spaces between words, which is a standard feature in many other languages. To address this, our system employs advanced tokenization techniques that are specifically designed for the Thai language.

The second step involves removing stop words and question marks from the queries. Stop words, which are commonly used words such as 'and', 'is', 'at', etc., typically provide little value for document retrieval and can unnecessarily complicate the process. Similarly, question marks are also eliminated as they do not contribute to identifying relevant documents.

Next, we apply normalization methods to the queries. These methods are crucial in dealing with the Thai language's characteristics, such as multiple valid spellings for the same word and the use of tone marks. Our normalization process eliminates duplicate spaces, removes dangling characters, and consolidates repeating vowels. These actions help standardize the queries and make them more amenable to the subsequent document retrieval process.

Finally, after these pre-processing steps, keywords are extracted from the user queries. These keywords serve as the cornerstone of the document retrieval process, helping to identify and rank the most relevant documents in response to the user's query. Overall, despite the complexities presented by user queries and the Thai language, our system is designed to effectively process and handle these challenges, thereby ensuring accurate and efficient document retrieval. The final representation of each user query is a set of keywords derived from the original query. To provide a clear illustration of this process, Table 2 presents examples of original user queries in Thai and their corresponding extracted keywords, with English translations provided in parentheses.

| Query | Keywords |
|---|---|
| ขอข้อมูลนิคมอุตสาหกรรมในเมียนมาร์หน่อยค่ะ | ['นิคมอุตสาหกรรม', 'ข้อมูล', 'เมียนมาร์' ] |
| (May I get information on industrial estates in Myanmar?) | ['industrial estates', 'information', 'Myanmar'] |
| ประชากรในเมียนมาร์มีเท่าไรคะ | ['ประชากร', 'เมียนมาร์' ] |
| (What is the population in Myanmar?) | ['population', 'Myanmar'] |
| เมืองหลวงของกัมพูชาชื่ออะไรเหรอ | ['เมืองหลวง', 'ชื่อ', 'กัมพูชา' ] |
| (What is the name of the capital city of Cambodia?) | ['capital city', 'name', 'Cambodia'] |

Table 2: The original queries and corresponding keywords

### 3.4 Ranking Module

The ranking module in this system is responsible for ranking the retrieved documents based on their relevance to the user's query, considering both the document heading and content for ranking purposes. To rank documents based on their headings, each keyword is first mapped to a pre-trained word embedding vector using a deep learning model. This enables the model to capture the semantic and contextual meaning of each keyword in the document representation. The final word embedding vector is then obtained by calculating the mean of all the keyword vectors. This approach allows the model to capture the overall meaning and context of the document's heading, providing a more comprehensive representation for accurate ranking. Using the same approach, the user query is also converted into a word embedding vector. For each document, the system calculates the cosine similarity score between the word embedding vectors of its heading and the user's query. These scores are then used to create a heading ranking array, which ranks the documents based on their similarity to the query.

For content ranking, the BM25 algorithm is employed, taking into account the frequency and importance of query terms in the document to rank the documents based on their relevance to the user's query. The BM25 algorithm considers both term frequency (TF) and inverse document frequency (IDF) to calculate the relevance score of a document to a query. The TF component measures the frequency of the query terms within a document, while the IDF component measures the rarity of the query terms in the document collection. By combining these two components, the BM25 algorithm can capture both the specificity of the query and the importance of the terms in the document.

The rankings obtained from both approaches are combined using the Borda count method. This method produces a final rank by assigning weights to the ranks of each document in both arrays. The document with the highest final rank is recommended as the most relevant document to the user's query. The Borda count method is used because it provides a more robust and accurate ranking by taking into account the rankings of all documents, rather than just focusing on the top-ranked documents. It is also known for its fairness, simplicity, and robustness, making it a reliable and widely used approach for rank aggregation in various applications (Dwork, Kumar, Naor, & Sivakumar, 2001; Renda & Straccia, 2003).

Algorithm 1 outlines the procedure for ranking and retrieving documents that are most relevant to a user's query. The algorithm starts by extracting a set of keywords, denoted as $K_{query}$ from the user's query. For each document $d$ in the database, the algorithm retrieves a set of heading keywords $K_{document}$ and a set of content tokens $C_{document}$ that have been preprocessed and stored.

Two metrics are calculated for ranking:

1. The BM25 score between $K_{query}$ and $K_{document}$, quantifies the relevance of the document's headings to the query.

2. The cosine similarity score between $K_{query}$ and $C_{document}$, which measures the angular distance between the keyword vectors, thereby indicating content relevance.

The Borda count method is then used to aggregate these scores into a final ranking score for each document. Once all the documents have been scored, they are sorted in descending order based on their final ranking scores. The top-ranked document is then returned to the user, completing the retrieval process.

---

**Algorithm 1** The proposed ranking algorithm

---

**Require:** User query $q$
**Ensure:** The top-ranked document

---

1: **Steps:**
2: Extract the set of keywords $K_{query}$ from $q$
3: **for** each document $d$ in the database **do**
4:    Retrieve the set of heading keywords $K_{document}$
5:    Retrieve the set of content tokens $C_{document}$
6:    Calculate the BM25 score between $K_{query}$ and $K_{document}$
7:    Calculate the cosine similarity score between $K_{query}$ and $C_{document}$
8:    Use the Borda count method to aggregate the BM25 score and cosine similarity scores arrays and calculate the final ranking score
9: **end for**
10: Sort the documents in descending order based on their final ranking scores
11: **return** The top-ranked document to the user

---

## 4 Results and Discussions

This section presents an evaluation of the proposed system's performance, including a comparison of different ranking approaches and an assessment of the system's effectiveness based on user survey feedback.

### 4.1 Performance Evaluation

This section embarks on a comprehensive evaluation and comparison of the proposed system's performance. Utilizing a test dataset of 406 matched query-document pairs, carefully constructed by Thai entrepreneurs and senior students in international business management, the system's effectiveness is thoroughly assessed. The documents, sourced from the Department of International Trade Promotion website, pertain to foreign trade and investment in four countries: Vietnam, Laos, Cambodia, and Myanmar. This collection of high-quality, up-to-date guidebooks is aimed at enhancing the competitiveness of Thai entrepreneurs in global markets

To validate the effectiveness of our proposed system, we undertook a detailed comparative analysis against two distinct ranking methods. The first of these methods based its ranking solely on documentcontent, employing the well-established BM25 method. Conversely, the second method leveraged only documentheadings, using the Word2Vec algorithm to calculate cosine similarity between query keywords and documentheadings.

The performance was evaluated with two primary metrics - accuracy rate and processing time in seconds. These measurements are outlined in Table 3. Through this analysis, our proposed methodology demonstrated superior performance, outstripping the other approaches in terms of accuracy. Our system attained an impressive accuracy rate of 65.76%. In comparison, the BM25 method, which strictly considers document content, yielded an accuracy of 57.88%. The Word2Vec-based method, focusing solely on document headings, achieved an accuracy of 62.81%. This clear advantage validates the effectiveness of our system's comprehensive approach, which blends the consideration of document content and headings to deliver more precise and relevant document suggestions.

However, it is worth noting that the proposed method took longer to execute, with an elapsed time of 0.034 seconds compared to 0.010 seconds for the BM25 method and 0.028 seconds for the Word2Vec-based method. Therefore, there is a trade-off between accuracy and speed. In real-world applications, the choice of ranking method will depend on the specific needs of the user. If accuracy is the top priority, then the proposed method may be the best choice, despite its longer execution time. It is also worth noting that the speed of the proposed method can be improved by adding more computing resources. This would allow for faster execution times without sacrificing accuracy. However, if speed is more important than accuracy, then the BM25 method or the Word2Vec-based method may be a better option, as they are faster but less accurate.

| Approach | Accuracy rate (%) | Computational time (seconds) |
|---|---|---|
| BM25 | 57.8818 | 0.0105 |
| Word2Vec-based | 62.8079 | 0.0279 |
| Proposed method | 65.7635 | 0.0344 |

Table 3: Performance comparison of ranking approaches

## 4.2 User Experience Evaluation

We utilized a user survey to ascertain the system's effectiveness from the user's perspective. Initially, participants were asked to submit several queries, with the system configured to suggest the top three most relevant documents in response to each query. Subsequently, participants were asked to respond to two key questions:

- On a scale of 1 - 5, with 1 signifying 'highly irrelevant' and 5 indicating 'highly relevant', how would you gauge the relevance of the documents suggested by the system in correspondence to your query?

- On a scale of 1 - 5, with 1 representing 'not useful at all' and 5 signifying 'extremely useful', how would you appraise the overall utility of the system in aiding with your query?

The intent behind the first question is to evaluate the user-perceived relevance of the suggested documents concerning their specific queries. Each participant was requested to rate the overall relevance of the recommended documents, with each rating corresponding to each query. The second question aims to measure the overall practical usefulness of the system from the users' perspective. This question is assessed on a per-user basis, providing an aggregate view of the system's utility across multiple queries made by each user. The user satisfaction survey involved 15 participants, composed of 10 entrepreneurs and five senior students specializing in international business management. Each participant submitted 3-4 queries, resulting in a total of 50 queries assessed by the system. The survey responses indicated a moderate level of satisfaction with the system's performance. The average rating for the first question, concerning the relevance of the documents suggested by the system, was 3.18 out of 5. The second question, evaluating the overall usefulness of the system, received a slightly higher average rating of 3.70 out of 5.

The survey results suggest that users generally found the system beneficial and moderately effective in assisting with their queries. An average rating of 3.70 out of 5 for the second question indicates that users see value in the system and deem it fairly useful in their query process. However, the average rating of 3.18 for the first question suggests that the system could potentially improve its document suggestion relevance. While this score is above average, it indicates some room for improvement. Users might have received some document recommendations that were not as relevant as they expected, or the relevance of the documents could have varied across different queries.

For example, one user queried "ขอข้อมูลเกี่ยวกับการนำเข้าสินค้าในเมียนมาร์หน่อยค่ะ" (Please provide information about importing goodsinto Myanmar). The relevant documents returned were as follows:

- กฎระเบียบการนำเข้าสินค้าในเมียนมาร์ (The regulations for importing goods in Myanmar)

- กฎหมายการลงทุนในเมียนมาร์ (Investment Laws in Myanmar)

- ระบบโลจิสติกส์และการขนส่งในเมียนมาร์ (Logistics and Transportation System in Myanmar)

The first result, "The regulations for importing goods in Myanmar," is the most relevant, directly addressing the specifics of import regulations. The second result, "Investment Laws in Myanmar," could provide a broader context about the legal environment for business and trade in Myanmar, which would also be useful for an entrepreneur considering importing goods into the country. The final result, "Logistics and Transportation System in Myanmar," could provide practical insights into the logistical aspects of importing goods, making it a valuable resource for understanding the transportation infrastructure and potential challenges.

While the search results for the query about importing goods into Myanmar are generally accurate and relevant, user satisfaction may vary depending on their specific information needs. The first result, "The regulations for importing goods in Myanmar," directly addresses the query, providing the most pertinent information about import regulations. However, the second and third results, though related to the broader context of conducting business in Myanmar, may not align precisely with the user's immediate information requirement about importing goods. Therefore, some users might feel these documents are not exactly what they were looking for, emphasizing the need for continual refinement in document suggestion relevance.

In conclusion, the results are promising, but there is room for improvement, particularly in enhancing the relevancy of the suggested documents. It might be worth considering refining the natural language processing techniques or information retrieval methods used in the system. Potential areas of focus could include improving keyword extraction, the ranking algorithm, or the approach to understanding the user's queries more accurately.

## 4.3 Discussion

Our proposed method's superior accuracy rate, surpassing both the BM25 and Word2Vec-based methods, can be attributed to its comprehensive approach that combines both content and heading-based strategies. By leveraging both these elements, our system is equipped to provide a more holistic understanding of the document's context, leading to more accurate document ranking. Additionally, our approach includes the novel feature of identifying country names in the queries, further refining search results by providing location-specific responses. This is a significant factor contributing to the enhanced accuracy of our method. However, this increased accuracy comes at the cost of execution speed. The comprehensive nature of our approach, integrating multiple techniques and additional features, inherently requires more computational time compared to simpler methods. This trade-off between accuracy and speed is a common challenge in information retrieval and natural language processing tasks.

In contrast, the BM25 method, which solely focuses on document content, performed less accurately but demonstrated a faster execution time. The BM25 algorithm, being a probabilistic approach, relies heavily on term frequency and inverse document frequency. While it is efficient, the method might overlook nuanced contextual elements present in the document headings, which could explain its lower accuracy rate. Similarly, the Word2Vec-based method, which only considers document headings, also exhibits a lower accuracy rate.

Despite using a sophisticated technique like Word2Vec, which understands semantic relationships between words, limiting the analysis to only headings might result in missing vital information from the main content of the document.

## 4.4    Limitations

While our system has shown effectiveness in assisting Thai entrepreneurs, there are some limitations. The system is primarily designed for Thai natural language queries, limiting its applicability to a more diverse, multilingual audience.  Additionally, the system relies heavily on guidebooks for its source material. These guidebooks cover a wide array of topics that may not all be relevant to a specific query. To address this, we segmented the guidebooks into smaller, topic-specific sections based on their headings. However, this segmentation itself could be a limitation, as it assumes that the headings accurately summarize the content, potentially leading to information being overlooked. Lastly, while the guidebooks are officially endorsed by Thai authorities, the system's reliance on them means that the timeliness of the information is contingent on these pre-existing resources. As such, regular updates are essential to maintain the system's relevance and effectiveness.

## 5    Conclusions

This paper presents a methodology that utilizes a ranking approach to provide a more accurate and relevant ranking of retrieved documents by considering both the document headings and content.  The system utilizes a deep learning model to measure the semantic similarity between a user's query and a document's heading. It also leverages the BM25 algorithm to rank documents based on their term frequencies in the content.  To retrieve the best-matched document, the system first preprocesses the documents and precomputes keyword vectors of all documents for later similarity calculation. Upon receiving a query, the system extracts a set of keywords from the query and compares it to heading keywords and content tokens of documents in the database. The system calculates the BM25 and cosine similarity scores, which are then aggregated using the Borda count method to produce a final ranking score for each document. The top-ranked document is returned as the most relevant document to the user.

The system was evaluated using a manually constructed dataset containing 406 queries. According to the experiment result, the proposed method achieved an accuracy rate of 65.76% with an elapsed time of 0.034 seconds. In comparison, the BM25 method achieved an accuracy rate of 57.88% with an elapsed time of 0.010 seconds, and the Word2Vec-based method achieved an accuracy rate of 62.81% with an elapsed time of 0.028 seconds. The proposed method outperformed both the BM25 method and the Word2Vec-based method in terms of accuracy rate.

The user satisfaction survey, involving 15 participants, provided insightful feedback on our system. Each participant submitted 3-4 queries, leading to a total of 50 queries assessed. The results suggested a moderate satisfaction level, with an average rating of 3.18 out of 5 for document relevance and a slightly higher average of 3.70 out of 5 for overall system usefulness. These ratings indicate that users found value in the system, deeming it beneficial for their query process. However, the feedback also highlights room for improvement, specifically in enhancing the relevance of suggested documents.

There are opportunities to enhance the matching module in the future to further improve the accuracy of the retrieved documents. Another potential area of improvement is to further enhance the document processing module to better capture the important information within the document content and headings. Additionally, exploring the use of more advanced deep learning models and incorporating automatic summarization techniques could also improve the accuracy of our approach.

## References

Amati, G., & Van Rijsbergen, C. J. (2002). Probabilistic models of information retrieval based on measuring the divergence from randomness. *ACM Transactions on Information Systems (TOIS)*, *20*(4), 357–389.

Dwork, C., Kumar, R., Naor, M., & Sivakumar, D. (2001). Rank aggregation methods for the web. In *Proceedings of the 10th international conference on world wide web* (pp. 613–622).

Galke, L., Saleh, A., & Scherp, A. (2017). Word embeddings for practical information retrieval. In *Informatik 2017* (pp. 2155–2167).

Guo, S., Alamudun, F., & Hammond, T. (2016). Résumatcher: A personalized résumé-job matching system. *Expert Systems with Applications*, *60*, 169–182.

Hiemstra, D. (2001). *Using language models for information retrieval*. Citeseer.

Huang, P.-S., He, X., Gao, J., Deng, L., Acero, A., & Heck, L. (2013). Learning deep structured semantic models for web search using clickthrough data. In *Proceedings of the 22nd acm international conference on information & knowledge management* (pp. 2333–2338).

Kawtrakul, A., Andres, F., Ono, K., Thumkanon, C., Thanyasiri, T., Khantonthong, N., . . . Buranapraphanont, N. (2000). The implementation of vlshds project for thai document retrieval. In *Proc. first international symposium on advance informatics, tokyo, japan.*

Lan, G., Ge, Y., & Kong, J. (2019). Research on scoring mechanism based on bm25f model. In *2019 ieee 8th joint international information technology and artificial intelligence conference (itaic)* (pp. 1782–1786).

Limkonchotiwat, P., Phatthiyaphaibun, W., Sarwar, R., Chuangsuwanich, E., & Nutanong, S. (2021). Handling cross and out-of-domain samples in thai word segmentation..

Lu, Z., & Li, H. (2013). A deep architecture for matching short texts. *Advances in neural information processing systems*, *26*.

Mitra, B., Craswell, N., et al. (2018). An introduction to neural information retrieval. *Foundations and Trends® in Information Retrieval*, *13*(1), 1–126.

Mitra, B., Diaz, F., & Craswell, N. (2017). Learning to match using local and distributed representationsof text for web search. In *Proceedings of the 26th international conference on world wide web* (pp. 1291–1299).

Mitra, B., Nalisnick, E., Craswell, N., & Caruana, R. (2016). A dual embedding space model for document ranking. *arXiv preprint arXiv:1602.01137*.

Naseri, S., Dalton, J., Yates, A., & Allan, J. (2021). Ceqe: Contextualized embeddings for query expansion. In *Advances in information retrieval: 43rd european conference on ir research, ecir 2021, virtual event, march 28–april 1, 2021, proceedings, part i 43* (pp. 467–482).

NAVER Corporation. (2020). *2020 NAVER annual report.* Retrieved from https://www.navercorp.com/en/investment/annualReport

Noraset, T., Lowphansirikul, L., & Tuarob, S. (2021). Wabiqa: A wikipedia-based thai question-answering system. *Information processing & management*, *58*(1), 102431.

Norvig, P. (2016). *How to write a spelling corrector.* http://norvig.com/spell-correct.html.

Phatthiyaphaibun, W. (2022). *Ltw2v: The large thai word2vec (1.0).* Retrieved from https://doi.org/10.5281/zenodo.7280277 doi: 10.5281/zenodo.7280277

Renda, M. E., & Straccia, U. (2003). Web metasearch: rank vs. score based rank aggregation methods. In *Proceedings of the 2003 acm symposium on applied computing* (pp. 841–846).

Robertson, S., Zaragoza, H., et al. (2009). The probabilistic relevance framework: Bm25 and beyond. *Foundations and Trends® in Information Retrieval*, *3*(4), 333–389.

Robertson, S., Zaragoza, H., & Taylor, M. (2004). Simple bm25 extension to multiple weighted fields. In *Proceedings of the thirteenth acm international conference on information and knowledge management* (pp. 42–49).

Soni, S., & Roberts, K. (2020). Patient cohort retrieval using transformer language models. In *Amia annual symposium proceedings* (Vol. 2020, p. 1150).

Sukhahuta, R., & Smith, D. (2000). Information extraction for thai documents. In *Proceedings of the fifth international workshop on on information retrieval with asian languages* (pp. 103–110).

Wu, H. C., Luk, R. W. P., Wong, K. F., & Kwok, K. L. (2008). Interpreting tf-idf term weights as making relevance decisions. *ACM Transactions on Information Systems (TOIS)*, *26*(3), 1–37.

Yang, W., Zhang, H., & Lin, J. (2019). Simple applications of bert for ad hoc document retrieval. *arXiv preprint arXiv:1903.10972*.

Zaragoza, H., Craswell, N., Taylor, M. J., Saria, S., & Robertson, S. E. (2004). Microsoft cambridge at trec 13: Web and hard tracks. In *Trec* (Vol. 4, pp. 1–1).

Zhai, C., & Lafferty, J. (2017). A study of smoothing methods for language models applied to ad hoc information retrieval. In *Acm sigir forum* (Vol. 51, pp. 268–276).

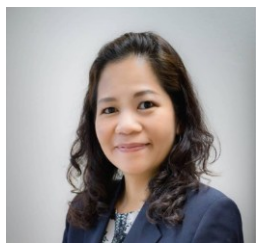

**Sirinda Palahan** is a faculty member in the Department of ICT at the School of Science and Technology, University of the Thai Chamber of Commerce, Thailand. With a Ph.D. in Computer Science and Engineering from Penn State University, USA, Dr.Sirinda Palahan possesses an academic background in the field. Her research interests primarily revolve around data analytics and natural language processing.